# The STAR W Spin Physics Program with $\sqrt{s}$ = 500 GeV Polarized pp Collisions at RHIC

W.W. Jacobs, for the STAR Collaboration

*Indiana University Cyclotron Facility and Department of Physics, Bloomington, Indiana, 47408 USA*
*jacobsw@indiana.edu*

**Abstract.** Production of *W* bosons in longitudinally polarized *pp* collisions provides an excellent tool to probe the flavor-dependence of sea quark polarizations in the polarized proton. Current status and future plans for the W physics program with the STAR detector at RHIC are presented along with remarks concerning our knowledge of the nature and origin of the partonic sea.

## INTRODUCTION

The *u* and *d* quark distributions in the proton are often described within the framework of SU(2), but no such symmetry governs the relative distributions of $\bar{u}$ and $\bar{d}$. Indeed, the latter relate to a fundamental physics question of nucleon structure: understanding the origin of the partonic sea and its dominant production mechanisms. If the production of sea quarks involves a purely perturbative mechanism (e.g., gluon splitting), one would expect equal numbers of quarks and anti-quarks, and quite similar contributions of all flavors of anti-quarks to the proton spin. In contrast, if the mechanism involves pion degrees of freedom (e.g., emission or absorption of Goldstone bosons at either the nucleon or quark level), one would expect $\bar{d} > \bar{u}$, from $u \rightarrow d\pi^+$ dominating $d \rightarrow u\pi^-$ transitions, and little contribution of the associated anti-quarks to the proton spin.

Experimental DIS results [1] some time ago indicated a significant violation of the Gottfried sum rule and hence evidence for flavor symmetry breaking in the light sea. More recently, a large imbalance in the proton sea was measured [2] via a fixed target Drell-Yan experiment (FNAL E866), where the extracted ratio $\bar{d}(x)/\bar{u}(x)$ rises to a maximum ~ 1.6 as a function of Bjorken x. The difference $\bar{d}-\bar{u}$ more readily exhibits the non-perturbative contributions and the data are in qualitative agreement with several models [3], including simple pion cloud models. More sophisticated chiral quark soliton approaches [3] incorporate gluon splitting as well as chiral origins of the sea quarks. Here the quarks are bound in a collective pion field in a manner consistent with a large-$N_c$ limit expansion approach to QCD. Polarized flavor asymmetries arise at a lower order in the expansion hierarchy, and indeed predict a large value for the difference $\Delta\bar{u}(x) - \Delta\bar{d}(x)$ of anti-quark polarizations [4].

Our current knowledge of polarized quark and anti-quark parton distribution functions comes mainly from global analyzes of DIS and SIDIS data, with recent addition of RHIC data constraining the gluons [5]. DIS measurements sum over flavor and are sensitive only to the combined contributions of quarks and anti-quarks; SIDIS can achieve separation via fragmentation function dependent analyzes. The global fitted polarized results for $\Delta \bar{u}(x)$ and $\Delta \bar{d}(x)$ have sizable uncertainties compared to the well determined quark + anti-quark sums. Nevertheless a positive $x(\Delta \bar{u}(x) - \Delta \bar{d}(x))$ is extracted, similar to $x(\bar{u}(x) - \bar{d}(x))$ but not as large as for the models discussed above.

## W PRODUCTION AS PROBE OF THE SEA

At the partonic level, $W^{-(+)}$ production is dominated by flavor dependent collisions whose leading order $\bar{u} + d$ ($\bar{d} + u$) are depicted in Fig. 1. The weak production processes involved are parity violating, yielding large longitudinal single spin asymmetries which can be measured as the difference between left-handed ($\sigma^-(W)$) and right-handed ($\sigma^+(W)$) production of W's divided by the sum:

$$A_L^W = \frac{1}{P} \frac{\sigma^+(W) - \sigma^-(W)}{\sigma^+(W) + \sigma^-(W)}. \tag{1}$$

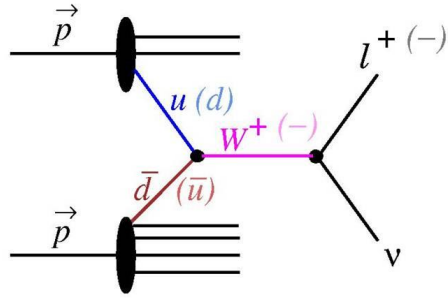

**FIGURE 1.** Partonic view of $W^{+(-)}$ production in pp collisions indicating $\bar{d} + u$ ($\bar{u} + d$) selectivity, respectively. The pure V-A interaction gives perfect spin separation: only left-handed quarks and right-handed anti-quarks couple to produce the left-handed W's.

Since the parity violation is maximal, the asymmetry $A_L^W$ is just the longitudinal polarization asymmetry of the contributing quark (anti-quark) in the proton:

$$\begin{aligned} A_L^{W^+} &\sim \bar{d}(x_1)\Delta u(x_2) - u(x_1)\Delta \bar{d}(x_2) \\ A_L^{W^-} &\sim \bar{u}(x_1)\Delta d(x_2) - d(x_1)\Delta \bar{u}(x_2) \end{aligned}, \tag{2}$$

where in general one has contributions from the two different cases of polarized quarks or anti-quarks as indicated in equation 2. From longitudinally polarized $pp$ collisions at RHIC, one can measure the asymmetry from either beam by helicity flip of one proton beam while averaging over the polarization of the other proton beam.

Production of the W's can be detected through their leptonic parity violating decays, $e^- + \bar{\nu}_e$ ($e^+ + \nu_e$), where only the charged lepton is measured. Discrimination of $+d$ ($\bar{d} + u$) quark combinations at STAR requires distinguishing between high $p_T$ $e^{-(+)}$ through their opposite charge sign, which in turn requires precise tracking information. Lepton kinematics most closely matches the W rapidity $y_W$ enabling a flavor separation of quark and anti-quark polarizations in far forward (backward) regions. There, $A_L^{W^-}$ approaches $\Delta d/d$ for $y_W \gg 0$; for $y_W \ll 0$ the asymmetry is $-\Delta \bar{u}/\bar{u}$ (for $W^+$ swap $u$ and $d$).

# STAR W PROGRAM

STAR is a large solid angle detector with full coverage by a Time Projection Chamber (TPC) and ElectroMagnetic Calorimetry (EMC). For the W program, tracking is required of high $p_T$ W decay $e^{+/-}$ leptons: at mid-rapidity, STAR will rely at first on the existing Time Projection Chamber (TPC). At forward rapidity, covered by the Endcap EMC (EEMC) and particularly favorable for clean signal extraction, new tracking capabilities will be provided by a Forward GEM Tracker (FGT) upgrade.

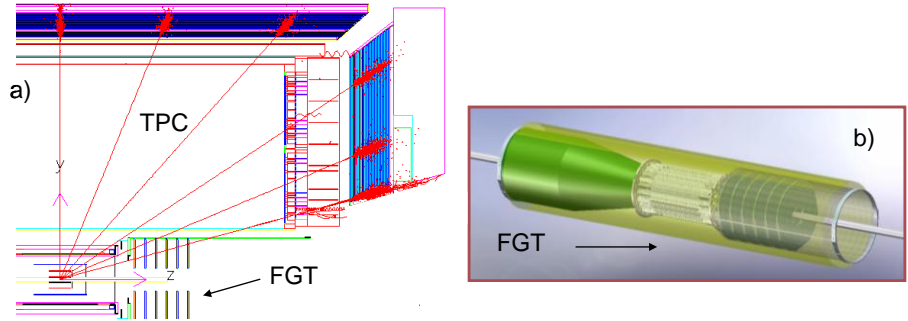

**FIGURE 2.** Forward Gem Tracker (FGT) upgrade. STAR quarter section a) with nominal trajectories indicating transition from TPC to FGT tracking. Central region b) with 6 triple GEM components

The FGT, currently under construction, will consist of six triple-GEM detectors as indicated in Fig 2, adding tracking hit points in the forward direction. The GEM disks, divided into quarter sections, have 2D readout (radius, azimuth) using APV25S1 chips and remote custom built DAQ readout. Square prototype GEM's have been tested successfully in FNAL beam. The addition of the FGT, including hit strips from the existing EEMC Shower Maximum detector (on the right in Fig. 2.a), is expected to achieve 80% charge separation out to the pseudorapidity limit $\eta \sim 2$ of the EEMC.

Extensive effort has been made recently to simulate electron–hadron discrimination in the forward direction covered by the EEMC. The analysis uses Pythia and GEANT simulations in the full STAR software framework. Special algorithms were used to pre-select events capable of depositing significant energy in the calorimeters, in order to generate sufficient hadronic background samples for the studies. These studies suggest that suppression of the QCD background compared to the *W* boson signal by several orders of magnitude can be accomplished by using the highly segmented STAR Electromagnetic Calorimeters, including isolation criteria suppressing jet events, and vetoing di-jet events based on the measured away side energy. The projected sensitivities, including errors arising from the simulated background studies, are shown by the on-axis data points in Fig. 3 for the EEMC ($1<\eta<2$) forward region, using an integrated *pp* collision luminosity of 300 pb$^{-1}$, average beam polarization 70% and with a W reconstruction efficiency of 70% as indicated from this analysis.

In Fig. 3 are also shown predictions using current results for the various underlying quark and anti-quark distributions. These are based on GRSV-STD , GRSV-VAL [6] and GS-A [7], calculated with RHICBOS [8], a program that incorporates an NLO calculation of spin observables in W production. GRSV-VAL is qualitatively similar to DSSV and uses a flavor asymmetric scenario for $\Delta \bar{u}$ and $\Delta \bar{d}$, while GRSV-STD is flavor symmetric. 300 pb$^{-1}$ provides significant discrimination between these models.

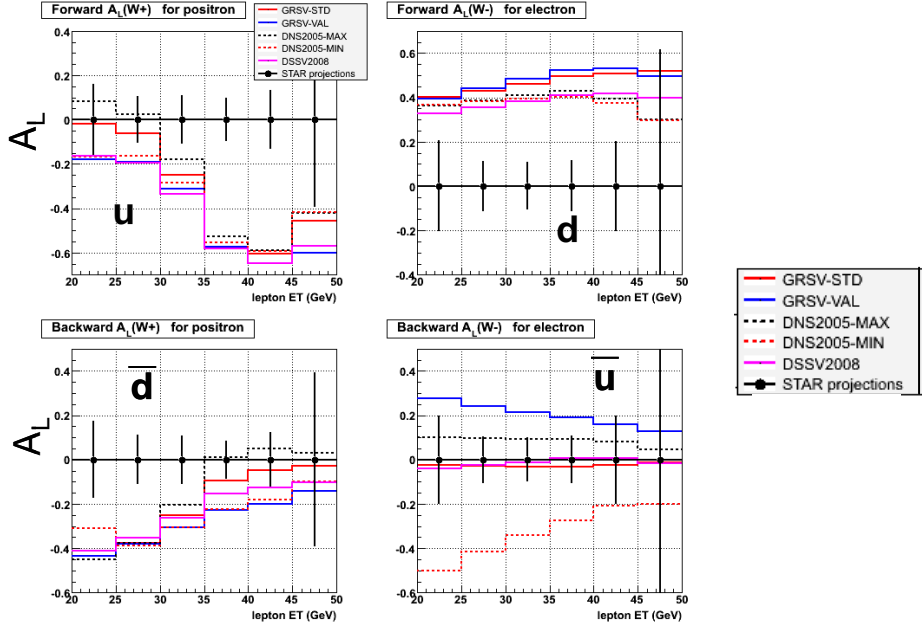

**FIGURE 3.** Estimation of precision for the single spin asymmetry for W production in pp collisions at $\sqrt{s}$ =500 GeV. Background simulations and their subtraction are used. Curves described in the text.

During RHIC run 9, the STAR experiment acquired its first *W* boson events in the mid-rapidity region from longitudinally polarized *p+p* collisions at $\sqrt{s}$ =500. Realistic simulations performed here also suggest S/B >1 is obtainable in the relevant $E_T$ range for the Barrel EMC; we expect to extract a mid-rapidity cross section from these data.

## SUMMARY AND OUTLOOK

The flavor and spin dependence of $\Delta\bar{q}$ ($\Delta q$) PDF's relate to fundamental questions concerning the origin of the sea and QCD. Parity Violating single spin asymmetries in W boson production provide one of the cleanest ways to probe this physics. At STAR an exciting program of W physics at $\sqrt{s}$ =500 GeV is underway. Full installation of a FGT forward tracking upgrade is anticipated prior to a long 500 GeV longitudinally polarized *pp* FY12 RHIC run. An integrated luminosity ~300 pb-1 with 70% polarized *p* in next ~5 years should yield significant results in both singular and global analyses.

## REFERENCES

1. P. Amaudruz *et al.*, Phys Rev. Lett. **66**, 2712 (1991); M. Arneodo *et al.*, Phys Rev D **50**, R1 (1994).
2. E. Hawker *et al.*, Phys. Rev. Lett. **80**, 3715 (1998); FNAL E866/NuSea, arXiv:hep-ex/0103030.
3. J.C. Peng *et al*. Phys. Rev. D **58**, 092004 (1998); P. Pobylitsa *et al.*, Phys. Rev D **59**, 034024 (1999).
4. B. Dressler *et al.*, Eur. Phys. J. **C14**, 147 (2000); *ibid,* **C18**, 719 (2001).
5. D. De Florian *et al.*, Phys. Rev Let.. **101**, 072001 (2008) ; *ibid*, arXiv:0904.382[hep-ph]
6. M. Gluck, E. Reya, M. Stratmann, and W. Vogelsang, Phys. Rev. D **63**, 094005 (2001).
7. T. Gehrmann and W.J. Sterling, Phys. Rev. D **53**, 6100 (1996).
8. P.M. Nadolsky and C.P. Yuan, Nucl. Phys. **B666**, 31 (2003).